\documentstyle[aps,prl,multicol,epsf]{revtex}

\begin{document}

\title{The Density Matrix Renormalization Group Method and
Large-Scale Nuclear Shell-Model Calculations}

\author{J. Dukelsky$^1$, S.~Pittel$^2$, S.S. Dimitrova$^3$, and M.V. Stoitsov$^3$ }

\address{$^1$ Instituto de Estructura de la Materia, Consejo
Superior de Investigaciones Cientificas, Serrano 123, 28006 Madrid, Spain \\
$^2$Bartol Research Institute, University of Delaware, Newark,
Delaware 19716, USA \\
$^3$Institute for Nuclear Research and Nuclear Energy, Bulgarian
Academy of Sciences, Sofia-1784, Bulgaria}
\maketitle
\vspace{7mm}
\begin{abstract} The particle-hole Density Matrix Renormalization Group
(p-h DMRG) method is discussed as a possible new approach to
large-scale nuclear shell-model calculations. Following a general
description of the method, we apply it to a class of problems
involving many identical nucleons constrained to move in a single
large j--shell and to interact via a pairing plus quadrupole
interaction. A single-particle term that splits the shell into
degenerate doublets is included so as to accommodate the physics
of a Fermi surface in the problem. We apply the p-h DMRG method to
this test problem for two $j$ values, one for which the shell
model can be solved exactly and one for which the size of the
hamiltonian is much too large for exact treatment. In the former
case, the method is able to reproduce the exact results for the
ground state energy, the energies of low--lying excited states,
and other observables with extreme precision. In the latter case,
the results exhibit rapid exponential convergence, suggesting the
great promise of this new methodology even for more realistic
nuclear systems. We also compare the results of the test
calculation with those from Hartree--Fock--Bogolyubov
approximation and address several other questions about the p-h
DMRG method of relevance to its usefulness when treating more
realistic nuclear systems.

\end{abstract}

\begin{center}
\end{center}

\begin{multicols}{2}

\section{Introduction}
The nuclear shell model \cite{R1} is one of the most powerful
approaches for a microscopic description of nuclear properties.
The low--energy structure of a given nucleus is described in this
approach by assuming an inert doubly--magic core and then
diagonalizing the effective hamiltonian within an active  valence
space consisting of at most a few major shells. Despite the
enormous truncation inherent in this approach, the shell-model
method as just described can still only be applied in very limited
nuclear regimes, namely for those nuclei with a sufficiently small
number of active nucleons or a relatively low degeneracy of the
valence shells that are retained. The largest calculations that
have been reported to date are for the binding energies of nuclei
in the $fp$--shell through $^{64}Zn$~\cite{R2}.

For heavier nuclei or nuclei farther from closed shells, one is
forced to make further truncations in order to reduce the number
of shell-model configurations to a manageable size. The most
promising approach now in use is to truncate on the basis of Monte
Carlo sampling\cite{R3}. In this way, it has recently proven
possible to extend the shell model beyond the $fp$--shell to
describe the transition from spherical to deformed nuclei in the
Barium isotopes~\cite{R4}.

Another attractive possibility is provided by the Density Matrix
Renormalization Group (DMRG), a method that was initially
developed and applied in the framework of low--dimensional quantum
lattice systems \cite{R5}. For a recent review, see ref.
\cite{R6}. A simplified version of this method was applied
\cite{R7} to the two-level pairing model, showing its convergence
properties. Subsequently, the method was extended to finite Fermi
systems and applied to a pairing problem of relevance to
ultrasmall superconducting grains \cite{R8}.   This new
methodology, which is referred to as the particle-hole (or p-h)
DMRG, was recently applied to a first test problem of relevance to
nuclear structure \cite{R9}. The application involved identical
nucleons moving in a single large-j shell under the influence of a
pairing plus quadrupole interaction with an additional
single-particle energy term that split the shell into
doubly-degenerate levels. Comparing with the results of exact
diagonalization, it was shown that the method leads to extremely
accurate results for the ground state and for low--lying excited
states without ever requiring the diagonalization of very large
matrices. Furthermore, even when the problem was not amenable to
exact solution, the method was seen to exhibit rapid exponential
convergence. All of this suggests that the DMRG method may indeed
be a practical means of carrying out reliable large--scale
shell-model calculations.

Before proceeding to realistic applications of the method, it is
important to clarify its various ingredients and assumptions.
Furthermore, it is critical to optimize how we apply the
methodology for the subsequent more realistic and complex
applications which will follow. With that in mind, we report here
the results of a more thorough investigation of the recent
application of the DMRG method to the one-orbit pairing plus
quadrupole model. Several of the questions that were hinted at in
ref. \cite{R9} are now addressed, including the feasibility of
applying the method to very much larger model spaces.

The outline of the paper is as follows. In Section II, we  review
the basic features of the DMRG method and provide a fairly
comprehensive description of the theoretical ingredients required
for its implementation. In Section III, we describe the one-shell
model that we will be using for the first tests of the DMRG method
in nuclear physics. In Section IV we present the results of these
test calculations for a problem of 10 identical nucleon restricted
to move in a single $j=25/2$ shell. This problem can be solved
exactly using the Lanczos algorithm, providing us with a good
testing ground of the methodology including several variants not
discussed in ref. \cite{R9}. In Section V, we discuss the
application of the method to a problem involving 40 nucleons in a
$j=99/2$ orbit, a problem well beyond the limits of exact
diagonalization. In Section VI, we summarize the principal
conclusions of this work and outline some future directions for
investigation.

\section{Overview of the p-h DMRG procedure}

\subsection{Qualitative description}

The basic idea of the p-h DMRG method is to {\it systematically}
take into account the physics of {\it all} single--particle
levels. The p-h nature of the method enters through a separation
of the active single-particle levels into two sets that are
separated by the Fermi surface; those above the Fermi surface are
called the particle levels and those below are called the hole
levels. The procedure begins by first taking into account the most
important levels, namely those that are nearest to the Fermi
surface, and then gradually including the others in subsequent
iterations. At each step of the procedure, a truncation is
implemented both in the space of particle states and in the space
of hole states, so as to optimally take into account the effect of
the most important states for each of these two subspaces of the
problem. The calculation is carried out as a function of the
number of particle and hole states that are maintained after each
iteration, with the assumption that these numbers are the same.
This parameter, which we call $p$, is gradually increased and the
results are plotted against it. Prior experience from other
applications of the methodology suggests that the results converge
exponentially with $p$. Thus, once we achieve changes with
increasing $p$ that are acceptably small we terminate the
calculation. Assuming that the convergence is rapid, the method
permits us to achieve an accurate description of the low-lying
states of the system without ever having to diagonalize the large
matrices that would arise in the absence of the DMRG truncation
strategy.

\subsection{A bit more detail}

It is useful to now put the above qualitative remarks about the
DMRG method into a more mathematical context.

The general problem that we wish to solve is a standard shell
model problem of $N_{\nu}$ neutrons and $N_{\pi}$ protons (total
$N=N_{\nu}+N_{\pi}$) interacting via a one- plus two-body
hamiltonian, with the neutrons and protons each restricted to an
active set of single-particle orbitals. For simplicity of
presentation, we will focus in this subsection on just one type of
particle, as is the case for the test applications we will report.
In a later subsection, we will note those modifications required
for neutron-proton systems.

We begin by splitting the set of multiply-degenerate
single-particle levels into an ordered set of doubly-degenerate
levels. This is most naturally done by carrying out an
axially-symmetric deformed Hartree Fock (HF) calculation for the
system. Each of the resulting HF levels is then doubly-degenerate.
Going from the set of spherical single-particle levels to the
axially-symmetric set of doubly-degenerate levels is schematically
illustrated in figure 1 for a problem involving 8 identical
nucleons in the $f-p$ shell.

\begin{figure}[htb]

\begin{center}
\leavevmode \epsfysize=7.5cm \epsffile{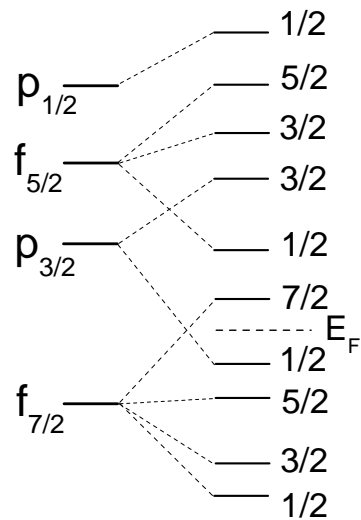}
\end{center}
\narrowtext\caption{\it Schematic illustration of the splitting of
the model-space single-particle levels into a set of
doubly-degenerate levels by an axially-deformed Hartree Fock
calculation. The dashed line represents the Fermi energy ($E_F$),
which separates the particle levels from the hole levels. Each
doubly-degenerate level is labelled by its angular momentum
projection on the intrinsic z-axis.} \label{fig1}
\end{figure}

Each of these doubly-degenerate single-particle levels admits
precisely four states. One is the state with no particles, one is
the state with one particle having positive $m$ projection, one is
the state with one particle having negative $m$ projection, and
the last is the state containing two particles, with both positive
and negative $m$.

This set of doubly-degenerate single-particle levels naturally
splits into two smaller sets, as likewise illustrated in figure 1.
Those levels above the (non-interacting) Fermi surface are the
particle levels and those below the Fermi surface are the hole
levels.  Reiterating what was stated earlier, we will
systematically take into account all of the levels of the problem,
by gradually moving away from the Fermi surface.

In this basis, the shell model hamiltonian takes the form
\begin{eqnarray}
H &=& \sum_{\alpha m} \epsilon_{\alpha m} a^{\dagger}_{\alpha m} a_{\alpha m}  \nonumber \\
&+& \frac{1}{4} \sum_{\alpha_1m_1\alpha_2m_2\alpha_3m_3\alpha_4m_4}
~ \langle \alpha_1m_1,\alpha_2m_2|V| \alpha_3m_3,\alpha_4m_4 \rangle  \nonumber \\
&& ~~~~~~~~~~~~~~~~~~~~~~~~~ \times a^{\dagger}_{\alpha_1m_1} a^{\dagger}_{\alpha_2m_2}
a_{\alpha_4m_4} a_{\alpha_3m_3} ~,
\label{ham}
\end{eqnarray}
with both a single-particle energy term and a two-body interaction.

At this point, let us assume that we have already treated some of
the levels of the problem, perhaps the first $n$ particle levels
above the Fermi energy (forming a {\em particle block}) and the
first $n$ hole levels below the Fermi energy (forming a {\em hole
block}). Let us also assume that we have precisely $p$ particle
states in the particle block and $p$ hole states in the hole
block, i.e. the same number for each.

Furthermore, we will assume that we have calculated and stored the
matrix elements of all possible sub-operators of the hamiltonian
within the $p$-dimensional particle and hole blocks, respectively.
For a one- plus two-body hamiltonian, the operators of relevance
are: $a^{\dagger}$, $a^{\dagger}a$, $a^{\dagger} a^{\dagger}$,
$a^{\dagger}a^{\dagger}a$ and $a^{\dagger} a^{\dagger} a a$, as
all others can be directly calculated from these using appropriate
hermitian adjoint relations.

We now  enlarge both the particle and hole blocks, by adding one
more (doubly-degenerate) level to each. Clearly the number of
states in the enlarged particle and hole blocks will be $4p$.
These states can be written as $|i,j \rangle = |i \rangle |j
\rangle$, where $i=1,...,p$ and $j=1,...,4$.

What we would like to do is to truncate both the particle and hole
blocks from these $4p$ states to the {\it optimum} $p$ states,
precisely the same number that we had before adding the new level.

How do we accomplish this?

The first step is to recalculate the matrix elements of all of the
hamiltonian sub-operators in the enlarged particle and hole
blocks. To show how this is done, we will focus on one specific
set of matrix elements, those of the operator
$a^{\dagger}_{\alpha_1m_1} a_{\alpha_2m_2}$. Its matrix elements
in the enlarged block can be readily expressed as

\begin{eqnarray}
\langle i, j | a^{\dagger}_{\alpha_1m_1} a_{\alpha_2m_2} | k,l
\rangle && =\langle i | a^{\dagger}_{\alpha_1m_1} a_{\alpha_2m_2}
| k \rangle ~\delta_{j,l}\nonumber \\
 &&+
\langle j | a^{\dagger}_{\alpha_1m_1} a_{\alpha_2m_2} | l \rangle ~\delta_{i,k} \nonumber \\
&&+ (-)^{n_k} \langle i | a^{\dagger}_{\alpha_1m_1} | k \rangle ~
\langle j | a_{\alpha_2m_2} | l \rangle \nonumber \\
&&- (-)^{n_k} \langle i | a_{\alpha_2m_2} | k \rangle ~ \langle
j | a^{\dagger}_{\alpha_1m_1} | l \rangle ~, \nonumber \\
\label{enlarge}
\end{eqnarray}
where $n_i$ is the number of particles in state $|i \rangle$.
Depending on which of the two subspaces the operator indices refer (either the
levels $1 \rightarrow n$ or the new level $n+1$), a different term in the sum applies.
For example, if both $\alpha_1,m_1$ and $\alpha_2,m_2$ refer to single-particle states
from the first $n$ levels, it is the first term that applies.

The key point is that all terms in (\ref{enlarge}) involve a
product of a matrix element from the previous $p\times p$ space
(levels $1 \rightarrow n$) and a matrix element for the level that
is being added. The matrix element from the previous space was
already calculated and stored, whereas the one from the new level
is very simple and can be written down analytically.

At this point we have calculated and stored all of the particle
and hole matrix elements in their enlarged blocks. The next step
is to construct the space of particle-hole states, by forming
products of states in the particle and hole blocks, viz: $|i_P,j_H
\rangle = |i_P \rangle |j_H \rangle$. This is usually referred to
as the superblock.

We only build states in the superblock with the correct number of
total particles $N$ for the problem of interest. That means that
we only consider states with the same number of particles and
holes. Furthermore, assuming that the hamiltonian is rotationally
invariant, it suffices to consider only those states for which the
total z-projection of the angular momentum is zero.

Within this product space, we then build the hamiltonian matrix. Using the simplified
notation $\gamma=\{\alpha m \}$, this can be expressed as

\begin{eqnarray}
\langle i_P,j_H | &H& | k_P,l_H \rangle = \sum_{\gamma}
\epsilon_{\gamma} \left[ \langle i_P | a^{\dagger}_{\gamma}
a_{\gamma} |k_P \rangle ~ \delta_{j_H,l_H} \right.\nonumber \\
 &~& ~~+~\left.\langle
j_H | a^{\dagger}_{\gamma} a_{\gamma} |l_H \rangle
~\delta_{i_P,k_P}\right] \nonumber \\
&~& ~~ +~\frac{1}{4} \sum_{\gamma_i,\gamma_j,\gamma_k,\gamma_l}
~\langle \gamma_i, \gamma_j | V | \gamma_k, \gamma_l \rangle \nonumber \\
&& ~\left[~~\delta_{i_P,k_P} ~
\langle j_H | a^{\dagger}_{\gamma_i} a^{\dagger}_{\gamma_j}
a_{\gamma_l} a_{\gamma_k} | l_H \rangle  \right.\nonumber \\
&& ~+~ 2  (-)^{n_{k_P}} \langle i_P | a^{\dagger}_{\gamma_i} | k_P \rangle
~ \langle j_H | a^{\dagger}_{\gamma_j} a_{\gamma_l} a_{\gamma_k}| l_H \rangle
\nonumber \\
&& ~+~ 2  (-)^{n_{k_P}} \langle i_P | a_{\gamma_l} | k_P \rangle ~
\langle j_H | a^{\dagger}_{\gamma_i} a^{\dagger}_{\gamma_j}
a_{\gamma_k}| l_H \rangle
\nonumber \\
&& ~+~  \langle i_P | a^{\dagger}_{\gamma_i}
a^{\dagger}_{\gamma_j} | k_P \rangle ~ \langle j_H |a_{\gamma_l}
a_{\gamma_k} |l_H\rangle
\nonumber \\
&& ~-~ 4 \langle i_P | a^{\dagger}_{\gamma_i} a_{\gamma_l} | k_P \rangle
~ \langle j_H |a^{\dagger}_{\gamma_j} a_{\gamma_k} |l_H \rangle
\nonumber \\
&& ~+~  \langle i_P |a_{\gamma_l} a_{\gamma_k}  | k_P \rangle
~ \langle j_H |a^{\dagger}_{\gamma_i} a^{\dagger}_{\gamma_j} |l_H\rangle
\nonumber \\
&& ~+~ 2  (-)^{n_{k_P}} \langle i_P | a^{\dagger}_{\gamma_j}
a_{\gamma_k} a_{\gamma_l}  | k_P \rangle ~ \langle j_H
|a^{\dagger}_{\gamma_i}| l_H \rangle
\nonumber \\
&& ~+~ 2  (-)^{n_{k_P}} \langle i_P | a^{\dagger}_{\gamma_i}
a^{\dagger}_{\gamma_j} a_{\gamma_l}  | k_P \rangle ~ \langle j_H
|a_{\gamma_k}| l_H \rangle
\nonumber \\
&& \left. ~~ \delta_{j_H,l_H} ~ \langle i_P |
a^{\dagger}_{\gamma_i} a^{\dagger}_{\gamma_j} a_{\gamma_l}
a_{\gamma_k} | k_H \rangle  \right]~. \label{superblockH}
\end{eqnarray}
Note that every term in eq. (\ref{superblockH}) involves a product of a matrix element in
the particle space and a matrix element of the {\it conjugate} operator in the hole space.
All of these matrix elements, for both particles and holes, have already been calculated and
stored in their respective enlarged blocks.

Next we diagonalize the superblock hamiltonian, viz:,
\begin{equation}
H|\Psi_k \rangle = E_k |\Psi_k \rangle ~,
\end{equation}
with
\begin{equation}
|\Psi_k \rangle = \sum_{i_P,j_P=1,4p} \Psi^{(k)}_{ij} |i_P \rangle
|j_H \rangle ~. \label{superblock}
\end{equation}
Note that both sums go over the $4p$ states in the respective
enlarged particle and hole blocks.

We now wish to construct the optimal approximation to the ground
state of the system (or to some low-lying set of states) that is
achieved when we only retain $p$ particle states and $p$ hole
states. By optimal, we will mean the approximation that maximizes
the overlap between the truncated state and the exact ground state
of the superblock ($|\Psi_1 \rangle$ in eq. (\ref{superblock})).
For now, we will ask for the optimal description of the ground
state; later we will discuss how to generalize this to several
states.

We will perform the optimized truncation in two stages, first
asking what is the optimal approximation when we truncate the
particle states and then asking what is the optimal approximation
when we truncate the hole states.  For simplicity of notation, we
will let $|\Psi \rangle=|\Psi_1 \rangle$ and $\Psi_{ij} =
\Psi^{(1)}_{ij}$.

Let us focus first on the particle block. We wish to find a subset
of particle states ($p$ in number), such that the truncated ground
state built up from these states has the largest possible overlap
with the exact superblock ground state. To accomplish this, we
first define the reduced density matrix for particles,

\begin{equation}
\rho^{P}_{i_Pi'_P} = \sum_{j_H=1,4p} \Psi_{i_Pj_H}
\Psi^*_{i'_Pj_H} ~.
\end{equation}
The reduced density matrix separates into blocks according to the
number of particles and the $m$ value.

We then diagonalize this $4p \times 4p$ matrix,
\begin{equation}
\rho^P |u^{\alpha}_P >  = \omega^P_{\alpha} | u^{\alpha}_P >  ~.
\end{equation}
A given eigenvalue $\omega^P_{\alpha}$ represents the probability
of finding the particle state $|u^{\alpha}_P >$ in the full ground
state wave function $|\Psi>$. Thus, the optimal truncation in the
sense described above corresponds to retaining the $p$
eigenvectors $|u^{\alpha}_P >$ that have the largest eigenvalues.

We then do exactly the same for the hole block. Namely, we
construct the reduced density matrix for holes,
\begin{equation}
\rho^{H}_{j_Hj'_H} = \sum_{i_P=1,4p} \Psi_{i_Pj_H}
\Psi^*_{i_Pj_H'} ~,
\end{equation}
diagonalize it, and then keep only the $p$ hole states with the
largest eigenvalues.

At this point, we have identified the $p$ particle states and $p$
hole states that best approximate the full ground state of the
coupled superblock. The final step before proceeding to the next
level is to transform all of the matrix elements that we
calculated in the $4p$-dimensional particle and hole spaces to the
optimal $p$-dimensional truncated spaces.

As noted earlier, we need not target our optimization to the
ground state only. We can target it to any set of states we wish.
If, for example, we wanted to optimally describe the lowest $L$
eigenstates of the system, we would build mixed density matrices

\begin{eqnarray}
\rho^P_{i_Pi'_P} &=& \frac{1}{L} \sum_{k=1,L} \sum_{j_H} \Psi^{(k)^*}_{i_Pj_H} \Psi^{(k)}_{i'_Pj_H} \nonumber \\
\rho^H_{j_Hj'_H} &=& \frac{1}{L} \sum_{k=1,L} \sum_{i_P}
\Psi^{(k)^*}_{i_Pj_H} \Psi^{(k)}_{i_Pj'_H}
\end{eqnarray}
and use them to choose the $p$ most important particle and hole
basis states to retain.

Once this series of steps has been implemented, we simply return
to the point where the discussion began and continue iteratively.
Namely we add the next levels, precisely as we did above, first
calculating and storing all matrix elements in the enlarged
particle and hole blocks, then constructing and diagonalizing the
hamiltonian matrix in the superblock, then constructing and
diagonalizing the reduced density matrices for particles and holes
(depending on the states we wish to target) and then truncating to
the most important $p$ states for particles and holes,
respectively, based on the eigenvalues of the associated density
matrices.

Were the number of particle and hole levels the same, we would
simply proceed as above to treat all levels. We would initialize
the iteration process by treating the two levels nearest to the
Fermi surface, whose matrix elements can be calculated very
simply. We would then continue to add levels as defined above.
Truncation would not be necessary until the number of
particle/hole states exceeds $p$. Prior to that point, it is not
necessary to construct the hamiltonian matrix in the superblock,
since it is only needed for the purpose of truncation. Note that
in the absence of truncation, the number of particle and hole
states after treating $n$ levels is $4^n$. Thus, until $4^n > p$,
truncation is not required.

When there are different numbers of particle levels and hole
levels, some modification to the above algorithm is needed. Assume
for example that there are more particle levels than hole levels,
so that the system is less than half full. We can continue to add
levels two at a time, one for particles and one for holes, until
we have exhausted all of the hole levels. From that point on,
however, there are no hole levels to add. Thus, in subsequent
iterations, we only add particle levels and only carry out the
optimized truncation for those states.

At each iteration, there are in fact $4p$ states for particles and
holes in the enlarged blocks required to build the superblock
hamiltonian. It is only after truncation that it is reduced to
$p$. Assuming that there are more particle levels than hole
levels, we have a choice of how to proceed after the hole levels
have been exhausted. One possibility is not to carry out a
truncation of the hole levels at its last iteration, keeping $4p$
hole states in subsequent iterations. The other is to carry out a
truncation at the last hole iteration and then keep only $p$ hole
states subsequently. The former is more accurate; the latter
reduces (perhaps significantly) the storage needs for the
calculation. Considering that storage issues are likely to be
especially important for subsequent applications of the
methodology to more complex systems, it is useful to assess the
accuracy lost if we keep only $p$ hole states after the hole
levels have been exhausted. This is one of the issues we will
address in the test applications to follow.

\subsection{Neutrons and Protons}

In the presence of both neutrons and protons, the same formalism
applies, but with a few minor practical modifications.

\begin{itemize}
\item
In a system of neutrons and protons, there are four distinct
blocks -- neutron particle, proton particle, neutron hole and proton hole.
\item
In a given iteration, we have a choice of how many and which
levels to include. One possibility is to add a particle level and
a hole level of the same type, switching between proton levels and
neutron levels from one iteration to the next. The other
possibility is to add four levels in an iteration -- one for each
of the four blocks. Both options can readily be implemented with
the formalism of the previous subsection and both permit us to
eventually include all active levels. We expect that the first
approach, in which we add a particle and a hole level in each
iteration, should suffice for heavy nuclei where the active
neutrons and protons occupy different major shells. For lighter
nuclei where they occupy the same major shells, it is most likely
preferable to add all four levels at the same time.
\item
The superblock basis in which the hamiltonian is diagonalized
consists of products of states of all four types -- neutron
particle, neutron hole, proton particle and proton hole. Only
those states with the correct numbers of neutrons and protons and
with total angular projection zero need be considered.
\item
The reduced density matrix for a given block requires a
contraction over the states of the other three blocks.
\item
We need to consider hamiltonians that include single-particle
energy terms for neutrons and protons as well as the
neutron-neutron, proton-proton and neutron-proton interactions.
The matrix elements for all of these terms can be readily obtained
from the same basic like-particle operators as before
($a^{\dagger}$, $a^{\dagger}a$, $a^{\dagger}a^{\dagger}$,
$a^{\dagger}a^{\dagger}a$ and $a^{\dagger}a^{\dagger}aa$).
\end{itemize}

\section{The model}

The first application of the p-h DMRG, as described in the
previous section, was to a problem of a very large number of
particles interacting via a pairing force and constrained to a set
of equally-spaced doubly-degenerate single-particle levels
\cite{R8}. This problem, which is of direct relevance to the
physics of ultrasmall superconducting grains, was solved at half
filling for up to 400 levels. Despite the enormity of the model
space, especially for a very large number of levels, the DMRG
method was able to very accurately reproduce the ground state of
the system, which could be obtained exactly using a method
pioneered by Richardson\cite{R10}. These results suggest that the
p-h DMRG method can very accurately describe the properties of
fermion systems that are dominated by a single collective degree
of freedom.

Nuclei, on the other hand, are characterized by several competing
collective degrees of freedom, most importantly those associated
with the pairing and quadrupole fields. Thus, before applying the
DMRG method with confidence to nuclear systems, it is necessary to
first demonstrate that it continues to work well in the presence
of such competing collective effects.

With that in mind, we report here a series of test calculations
for systems that admit pairing plus quadrupole correlations on the
same footing. The simplest such system is one involving identical
nucleons restricted to a single level with large degeneracy and
interacting via a pairing plus quadruple force. Unfortunately, the
($0^+$) ground state for an even number of particles occupying a
single level invariably has equal population of all $m$ substates
and thus such a problem does not admit a Fermi surface. Since a
Fermi surface is an important characteristic of real Fermi
systems, including nuclei, it is important to permit one in the
test calculations. Thus, in the calculations to follow we
supplement the pairing plus quadrupole interaction of our test
model with a single-particle energy term that splits the
multiply-degenerate large-j oprbit into a series of
doubly-degenerate levels. The hamiltonian we use takes the form

\begin{equation}
 H= - \chi Q \cdot Q - g  P^{\dagger}~ P -
\epsilon \sum_{m} |m| ~ a^\dagger_{jm} a_{jm} ~. \label{H}
\end{equation}
The last term splits the levels of the single-j shell into a set
of equally spaced levels of ``oblate" character, with the largest
$|m|$ value lowest. Because of the last term, the hamiltonian is
not in general rotationally invariant and thus its eigenstates do
not have conserved angular momentum. Rotational invariance is of
course recovered for $\epsilon=0$.

In the calculations we will describe, we use a scaled version of
the quadrupole-quadrupole strength $\chi$, defined according to
\begin{equation}
\chi = \frac{\langle j ||Q|| j \rangle ^2 }{2j+1} ~ \tilde{\chi}
~,
\end{equation}
where $\tilde{\chi}$ is the usual $Q \cdot Q$ coupling strength.
With this definition, $\chi$ and $g$ have the same dimensions,
which we will subsequently ignore.

\section{Results for 10 particles in a j=25/2 orbit}

\subsection{General remarks}

In this section, we report the results of calculations for a
system of 10 identical particles in a $j=25/2$ orbit. In this
case, there are 13 active doubly-degenerate orbits, 5 for holes
and 8 for particles. The Fermi surface lies between the $m=17/2$
hole level and the $m=15/2$ particle. For this problem, the size
of the space that would be required for exact diagonalization
(assuming that we only consider $m=0$ states) is 109,583, well
within the limits of standard diagonalization routines.  Thus, for
this set of calculations, we can readily compare the results of
calculations based on the DMRG methodology with the corresponding
exact results, and thereby obtain important insight into the
usefulness of the method. We will first report calculations for
the {\em standard scenario}, in which ({\em i}) we only target the
ground state in the DMRG truncation strategy, and ({\em ii}) we
include the full complement of $4p$ hole states in the hole block
after all hole levels have been treated. We will then present
results for scenarios that systematically modify these two
assumptions.

\subsection{Results for standard scenario}
Here we present results in which we only target the ground state
and we include all $4p$ hole states in the hole block after the
hole levels have been exhausted.

In our earlier paper, we presented results for two cases, one with
$\chi=1,~g=0,~\epsilon=0.1$ and one with
$\chi=1,~g=0.5,~\epsilon=0.1$. In both cases, we achieved accuracy
for the ground state energy of roughly $1$ part in $10^6$ with
reasonable values of $p$.

Obviously the case with $g=0$ involves pure quadrupole
correlations. On the basis of HFB calculations that we have now
performed, we conclude that the $g=0.5$ case likewise is dominated
by quadrupole correlations, with pairing correlations minimal.
Since it is our desire here to assess the DMRG method in the
typical nuclear scenario in which different correlation effects
compete, we will show throughout this section results for a
further-enhanced pairing interaction, namely for
$\chi=1,~g=0.1,~\epsilon=0.1$. In this case, the HFB calculations
show a well developed superconducting solution. This can be
readily seen by noting that for this problem the ratio of the
average pairing gap to the mean spacing between HF levels is
$0.39/0.18=2.1$.

Figure 2 shows results for the DMRG correlation energies, defined
as the gain in energy relative to Hartree Fock approximation.  The
exact and HFB values are

\begin{equation}
E_{corr}^{Exact}= -0.70633 ~~~;~~~ E_{corr}^{HFB} =-0.20641
\end{equation}

\begin{figure}[htb]
\begin{center}
\leavevmode \epsfxsize=8cm \epsffile{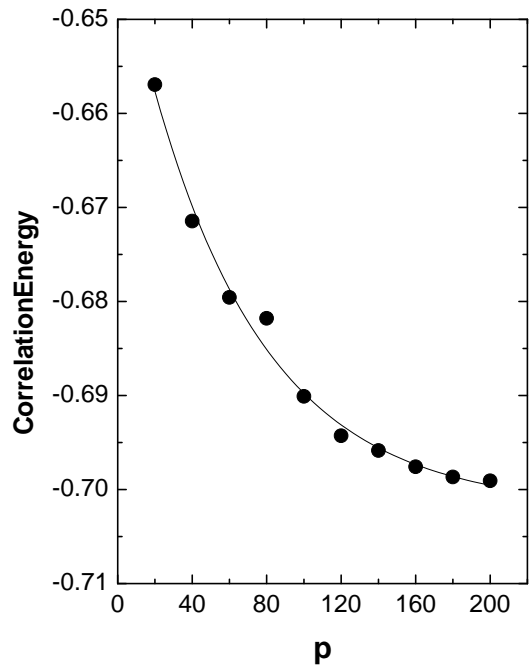}
\end{center}
\narrowtext\caption{\it DMRG correlation energies as a function of
$p$ for a system of 10 particles in a $j=25/2$ orbit subject to a
hamiltonian with $\chi=1$, $g=0.1$, and $\epsilon=0.1$. The solid
line represents an exponential fit to the DMRG results. }
\label{fig2}
\end{figure}
\noindent By a $p$ value of $200$, we obtain more than 99\% of the
exact correlation energy, a dramatic improvement over HFB
approximation which only gets 28\%.

For $p=200$ the largest superblock matrix that had to be
diagonalized had a dimension of $2,886$. For $p=40$, the maximum
dimension was $232$; nevertheless, for this value of $p$ we still
achieve over 95\% of the full correlation energy.

Table 1 shows results for the excitation energies of the lowest
states of this same system. Despite the fact that we only targeted
the ground state in the density matrix phase of these
calculations, the agreement is as good for the excitation energies
of these low-lying states as it is for the ground state energy. By
$p=200$, all three excitation energies are reproduced to 1\% or
better.

In Figure 3, we show the exact, DMRG and HFB results for the
occupation numbers of the particle and hole levels associated with
the ground state solution. The DMRG results, which are shown for
$p=200$, are in excellent agreement with the exact results for all
levels with appreciable occupation. In contrast, the HFB
calculation does not obtain sufficient depletion of the Fermi sea.

\begin{figure}[htb]
\begin{center}
\leavevmode \epsfxsize=8cm \epsffile{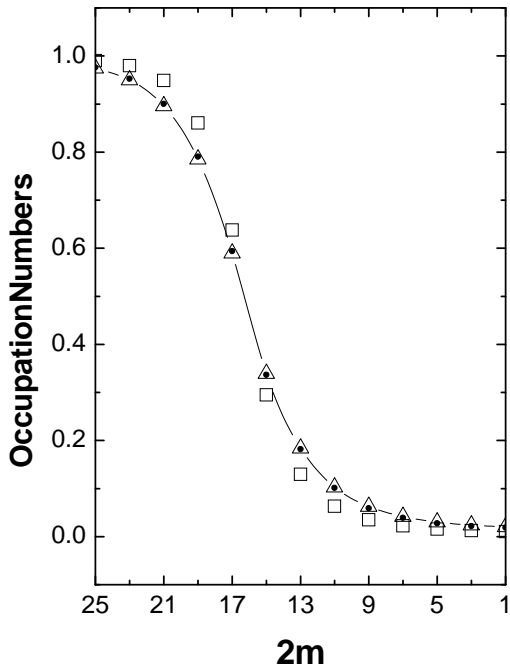}
\end{center}
\narrowtext\caption{\it The occupation numbers for the same system
as in figure 2. The horizontal label $2m$ is twice the
z-projection of the angular momentum. The solid circles (connected
to guide the eye by the solid line) represent the exact results;
the open diamonds represent the DMRG results for $p=200$; the open
squares represent the HFB results. } \label{fig3}
\end{figure}

From these results, we conclude that the p-h DMRG method is  able
to describe with extreme accuracy the low-lying properties of
complex many-body systems with competing collective features.

\subsection{Dependence on the number of states targeted}

What is the importance of including excited states in the
targeting procedure? We address this in Figure 4, where we compare
the results for the ground state energy and the energies of
low-lying excited states when different sets of states are
targeted. The solid triangles refer to results when only the
ground state is targeted ($L=1$). The open circles are the results
obtained when the lowest four states are all targeted in the
optimization procedure ($L=4$). We also include in the figure the
exact results.

\begin{figure}[htb]
\begin{center}
\leavevmode \epsfxsize=9cm \epsffile{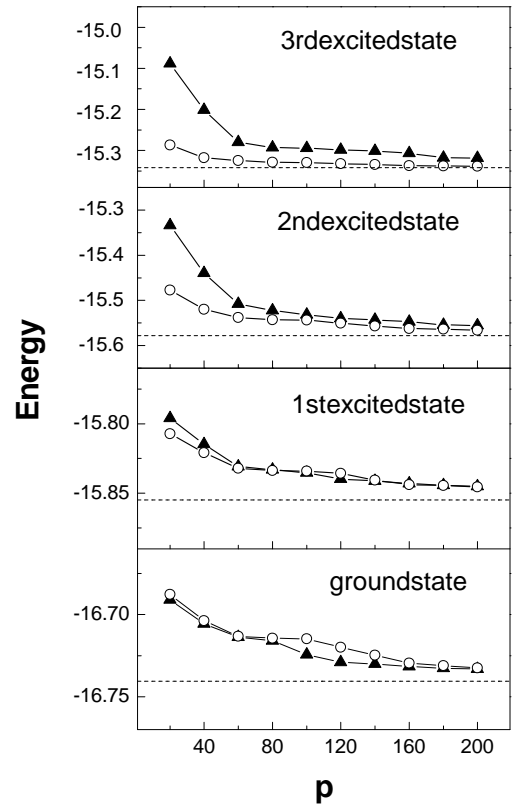}
\end{center}
\narrowtext\caption{\it The energy of the ground state and of the
three lowest excited states for the same system as in figure 2.
The solid triangles are the results when only the ground state is
targeted in the optimization procedure; the open circles refer to
calculations when the lowest four states of the system are
targeted simultaneously. The exact results are represented by
dashed lines. } \label{fig4}
\end{figure}

The results are not unexpected. The ground state is better
described when it is the only state targeted in the optimization
procedure. Furthermore, in such a scenario a reasonable
description of low-lying excited states is achieved. When several
states are targeted, the agreement gets worse, albeit only
marginally, for the ground state, but becomes better for the
excited states. What is most important, however, is that once $p$
becomes large enough, there is no discernable difference between
the two sets of results, both of which are also in excellent
agreement with the exact eigen-energies.

\subsection{Dependence on how we treat the hole block after all
hole levels have been exhausted}

Next we address the effect of maintaining all $4p$ hole states
following treatment of the last hole level. This was addressed by
carrying out the DMRG calculations in two ways, namely  keeping
only $p$ hole states after the last hole iteration or keeping
$4p$. For a value of $p=200$, we find for the ground state energy
a value of $-16.7331$ when only $p$ states are retained and a
value of $-16.7333$ when all $4p$ are kept. The gain in energy
through the improved treatment of the hole states is just $1$ part
in $10^6$. On the basis of these results, the subsequent
calculations we describe will be based on the simplifying, but
highly justified, assumption that we maintain just $p$ hole states
in those iterations that only add particle levels.

\section{Results for 40 particles in a j=99/2 orbit}

The excellent quality of the results obtained in the test
calculations of the previous section has encouraged us to treat a
much larger system, one for which exact diagonalization is not
possible. Following an optimization of the DMRG calculational
methodology, we are now able to treat systems significantly larger
than were feasible in ref. \cite{R9}. Here we report the largest
calculation we have so far carried out -- for a system of 40
particles occupying  a $j=99/2$ orbit. In this case, the exact
calculation would involve a hamiltonian matrix of dimension $3.84
\times 10^{25}$, obviously much too large to treat without
dramatic truncation. The parameters of the hamiltonian of eq.
(\ref{H}) that were used in this calculation are: $\chi=1, \;
g=0.1$ and $ \epsilon=0.2$. The calculations targeted only the
ground state and retained only $p$ hole states after treating the
$20$ hole levels. [The latter is especially important when
treating extremely large problems such as this one.] The results
for the ground state correlation energy are shown in figure 7. By
$p=100$, we obtain a ground state correlation energy of
$-\,2.8994$. An exponential fit to the DMRG results, likewise
indicated in the figure, give an asymptotic result for the ground
state energy of $-\,2.8990 \pm 0.0003$. This is close enough to
our DMRG result with $p=100$ to suggest that we have achieved
accuracy for the ground state correlation energy to 1 part in
$10^4$.

We have also carried out an HFB calculation for this system, which
confirms that the ground state is superconducting. The correlation
energy achieved in this calculation is $-\,1.7902$, less than 62\%
of that estimated from the exponential fit to the DMRG results.

\begin{figure}[htb]
\begin{center}
\leavevmode \epsfysize=8cm \epsfxsize=7cm \epsffile{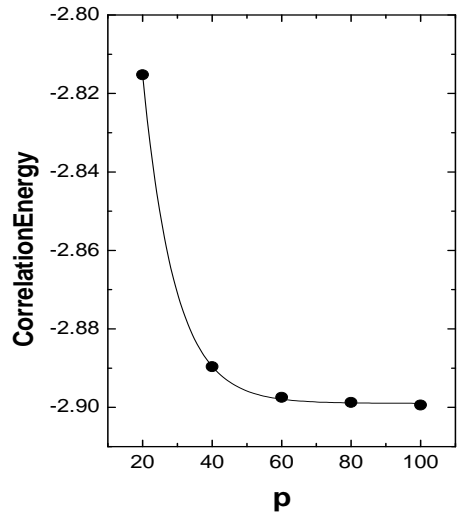}
\end{center}
\narrowtext\caption{\it The ground-state energy for a system of 40
identical nucleons in a $j=99/2$ orbit with the hamiltonian
parameters $\chi=1, \; g=0.1$ and $ \epsilon=0.2$. An exponential
fit to these results is also plotted.} \label{fig5}
\end{figure}

\section{Conclusion}

In this paper, we describe the recently-developed particle-hole
Density Matrix Renormalization Group method and report its test
application to a problem involving many identical nucleons
constrained to a single-j shell and subject to a hamiltonian with
a pairing plus quadrupole interaction and a single-particle term
that splits the shell into doublets. Earlier work along these
lines was reported in ref. \cite{R9}. Here we develop the
formalism in greater detail and report calculations that address
some important aspects of the method that needed to be clarified
before the method could be meaningfully applied to more realistic
nuclear systems.

In our view, the current calculations confirm and expand on the
conclusions reported in ref. \cite{R9} regarding the use of the
method in large-scale shell-model calculations. As long as there
is a well defined Fermi surface in the problem, the method is able
to produce extremely accurate results for the ground state energy
of the system and for the energies of low--lying excited states as
well, even in the presence of competing collective correlation
effects. It is also able to accurately reproduce other
(number-conserving) properties of the system, including for
example  occupation numbers. Based on an optimized computational
methodology, we are now able to treat much larger problems than in
ref. \cite{R9}, and this bodes well for subsequent more-realistic
applications of the methodology. In all cases we have studied,
even those very large-scale problems we are now able to handle
only because of the optimized methodology, we seem to be able to
obtain accurate results while diagonalizing matrices of moderate
dimensions. Critical to the success of the method is the rapid
exponential convergence that it typically exhibits as a function
of the number of particle and hole states maintained in a given
calculation.

Clearly the next step is for us to further develop the method for
use in realistic nuclear systems. This will mean including both
neutrons and protons, several non-generate single-particle levels,
and general nuclear hamiltonians. The framework for these
extensions is spelled out in Section II and is currently in the
process of being implemented.

\acknowledgments {This work was supported in part by the National
Science Foundation under grant \# PHY-9970749, by the Spanish DGI
under grant BFM2000-1320-C02-02,  by NATO under grant
PST.CLG.977000, by the Bulgarian Science Foundation under contract
$\Phi-905$ and by the Bulgarian--Spanish Exchange Program under
grant \# 2001BG0009. One of the authors (SDD) would also like to
acknowledge the partial support of a Fulbright Visiting Scholar
Grant}

\begin{table}[htb]
\begin{center}
{\bf Table 1:} Excitation energies for 10 particles in a $j=25/2$
level. The Hamiltonian parameters are: $\chi=1$, $g=0.1$ and
$\epsilon=0.1$. \\
\end{center}
\begin{tabular}{lcccc}
& $p$ & $E_1$ & $E_2$ & $E_3$\\
\tableline
& 40& 0.89040 & 1.26036  & 1.49354\\
& 80 & 0.88266 & 1.18608& 1.41182\\
& 120 & 0.88791 & 1.18077 & 1.41795\\
& 160 & 0.88794 & 1.17722 & 1.41456\\
& 200 & 0.88784 & 1.17302 & 1.41024\\
\hline & $Exact$ & 0.88578 &1.16245  &1.39915
\end{tabular}
\end{table}

\end{multicols}

\end{document}